\documentclass[12pt]{article}
\usepackage{epsfig}
\usepackage{lscape}
\usepackage{a4p}
\usepackage{cite}

\newcommand{\ks}{\mbox{K}^0_{\mathrm S}}

\newcommand{\sqee}{\sqrt{s_{\rm ee}}}

\def\ee{\mbox{e}^+\mbox{e}^-}

\def\pT{p_{\rm T}}
\def\pt{p_{\rm T}}

\def\qqbar{\mbox{q}\overline{\mbox{q}}}
\def\dspt{{\rm d}\sigma/{\rm d}p_{\rm T}} 
\def\dseta{{\rm d}\sigma/{\rm d}|\eta|}

\begin{document}

\renewcommand{\thefootnote}{\arabic{footnote}}


\begin{titlepage}
\begin{center}{\large   EUROPEAN ORGANIZATION FOR NUCLEAR RESEARCH
}\end{center}\bigskip
\begin{center}
\begin{flushright}
CERN-PH-EP/2006-038\\
18th October 2006\\

\end{flushright}
\vspace{2.5cm}
\huge{\bf \boldmath Inclusive Production of Charged Hadrons
in Photon-Photon Collisions \unboldmath} \\
\end{center}
\vspace{1.5cm}
\begin{center}
\LARGE{The OPAL Collaboration}\\
\end{center}

\vspace{1cm}

\begin{abstract}
  The inclusive production of charged hadrons in the collisions of
  quasi-real photons ($\ee \rightarrow\ee$+X) has been measured using
  the OPAL detector at LEP.  The data were taken at $\ee$
  centre-of-mass energies from $183$ to $209$~GeV.  The differential
  cross-sections as a function of the transverse momentum and the
  pseudorapidity of the hadrons are compared to theoretical
  calculations of up to next-to-leading order (NLO) in the strong coupling
  constant $\alpha_{s}$. The data are also compared to a
  measurement by the L3 Collaboration, in which a large deviation from
  the NLO predictions is observed.
\end{abstract}

\vspace{2.5cm}

\begin{center}{\large
(Submitted to Physics Letters B)
}\end{center}

\end{titlepage}

\begin{center}{\Large        The OPAL Collaboration
}\end{center}\bigskip
\begin{center}{
G.\thinspace Abbiendi$^{  2}$,
C.\thinspace Ainsley$^{  5}$,
P.F.\thinspace {\AA}kesson$^{  7}$,
G.\thinspace Alexander$^{ 21}$,
G.\thinspace Anagnostou$^{  1}$,
K.J.\thinspace Anderson$^{  8}$,
S.\thinspace Asai$^{ 22}$,
D.\thinspace Axen$^{ 26}$,
I.\thinspace Bailey$^{ 25}$,
E.\thinspace Barberio$^{  7,   p}$
% ...
}\end{center}\bigskip
\bigskip
$^{  1}$School of Physics and Astronomy, University of Birmingham,
Birmingham B15 2TT, UK
\newline
$^{  2}$Dipartimento di Fisica dell' Universit\`a di Bologna and INFN,
I-40126 Bologna, Italy
\newline
% ...
$^{  5}$Cavendish Laboratory, Cambridge CB3 0HE, UK
\newline
% ...
$^{  7}$CERN, European Organisation for Nuclear Research,
CH-1211 Geneva 23, Switzerland
\newline
$^{  8}$Enrico Fermi Institute and Department of Physics,
University of Chicago, Chicago IL 60637, USA
\newline
% ...
$^{ 21}$Department of Physics and Astronomy, Tel Aviv University,
Tel Aviv 69978, Israel
\newline
$^{ 22}$International Centre for Elementary Particle Physics and
Department of Physics, University of Tokyo, Tokyo 113-0033, and
Kobe University, Kobe 657-8501, Japan
\newline
% ...
$^{ 25}$University of Victoria, Department of Physics, P O Box 3055,
Victoria BC V8W 3P6, Canada
\newline
$^{ 26}$University of British Columbia, Department of Physics,
Vancouver BC V6T 1Z1, Canada
\newline
% ...
$^{  p}$ now at The University of Melbourne, Victoria, Australia
\newline
% ...

\newpage
\section{Introduction}

Hadronic interactions of two photons lead to the production of hadrons
whose properties in Quantum Chromodynamics (QCD) depend on the
underlying partonic processes, as well as on the way in which the
partons are transformed into observable hadrons. The study of
inclusive hadron production is therefore an appropriate tool to
investigate the validity of QCD in hadronic photon-photon
interactions. Furthermore a drastic discrepancy between data and NLO
QCD has recently been observed in this process~\cite{bib-l3} which
would indicate the breakdown of this otherwise so successful theory
and hence requires further study. Our measurements complement similar
studies of jet production~\cite{bib-opaljets, bib-l3jets}. The same
partonic processes are at work in QCD in both cases, but hadronisation
is usually treated differently. To predict the production of single
hadrons, universal fragmentation functions obtained from fits to
independent data are utilised to describe the transition from partons
to hadrons.  Jet observables on the other hand are designed to keep
the influence of hadronisation processes small. The remaining
distortion with respect to partonic jet observables is estimated using
hadronisation models.  The reliable calculation of an observable in
perturbative QCD requires a sufficiently high energy scale which in
our case is provided by the transverse momenta of the charged
particles. First results have been published by OPAL at $\ee$
centre-of-mass energies $\sqee=161$ and $172$~GeV
~\cite{bib-opalhadr}.  L3 has published results in the range
$\sqee=189-202$~GeV ~\cite{bib-l3}. This paper extends the OPAL
measurements of the transverse momentum spectra and
pseudorapidity\footnote{In the OPAL coordinate system the $x$ axis
  points towards the centre of the LEP ring, the $y$ axis points
  upwards and the $z$ axis points in the direction of the electron
  beam.  The polar angle $\theta$, the azimuthal angle $\phi$ and the
  radius $r$ denote the usual spherical coordinates. The
  pseudorapidity {$\eta$} is defined as $\eta=-\ln\tan(\theta/2)$. }
distributions of charged hadrons in~\cite{bib-opalhadr} using data
taken at $\ee$ centre-of-mass energies from $\sqee=183$ to $209$~GeV,
representing a roughly thirty-fold increase in integrated luminosity.

The interactions of the photons can be modelled by assuming that each
photon can either interact directly or appear resolved through its
fluctuations into hadronic states.  In leading order QCD this model
leads to three different event classes for $\gamma\gamma$
interactions: direct, single-resolved and double-resolved, where
resolved means that the partons (quarks or gluons) inside the hadronic
photon take part in the hard interaction.  The probability to find
partons in the photon is parametrised by parton density functions.

At LEP the photons are radiated by the beam
electrons\footnote{Positrons are also referred to as electrons.} and
carry mostly small negative four-momenta squared, $Q^2$.  In this paper events
are considered only if the electrons are scattered at small angles and
are not detected.  Both photons are therefore quasi-real ($Q^2 \approx
0$~GeV$^2$). Differential hadron production cross-sections are
measured as a function of the transverse momentum and the
pseudorapidity of charged hadrons. The inclusive cross-sections are
compared to Monte Carlo (MC) event generators in leading order
$\alpha_s$ and theoretical calculations in next-to-leading order (NLO)
$\alpha_s$ for this process.


\section{The OPAL detector}

A detailed description of the OPAL detector can be found
elsewhere~\cite{opaltechnicalpaper}.  The central tracking was
performed inside a solenoidal magnet which provided a uniform axial
magnetic field of 0.435~T along the beam axis. Starting with the
innermost components, the tracking system consisted of a high
precision silicon microvertex detector, a precision vertex drift
chamber, a large volume jet chamber with 159 layers of axial anode
wires and a set of $z$ chambers measuring the track coordinates along
the beam direction.  The transverse momenta, $\pt$, of tracks are
measured with a precision parametrised by
$\sigma_{\pt}/\pt=\sqrt{0.02^2+(0.0015\cdot \pt)^2}$ ($\pt$ in GeV) in
the central region $|\cos\theta|<0.73$. In this paper transverse is
always defined with respect to the beam axis.

The magnet was surrounded in the barrel region ($|\cos\theta|<0.82$)
by a lead glass electromagnetic calorimeter (ECAL) and a hadronic
sampling calorimeter (HCAL), which in turn were surrounded by muon
chambers. Similar layers of detectors were installed in the endcaps
($0.82<|\cos\theta|<0.98$).  The small angle region from 47 to
140~mrad around the beam pipe on both sides of the interaction point
was covered by the forward calorimeters (FD) and the region from 33 to
59~mrad by the silicon tungsten luminometers (SW). The latter were used
to determine the luminosity by counting small-angle Bhabha scattering
events.


\section{Kinematics and Monte Carlo simulation}
\label{sect-kin}

The properties of the two interacting photons ($i=1,2$) are described
by their negative four-momentum transfers
$Q_{i}^2$ and their invariant mass.  Each $Q_i^2$
is related to the electron scattering angle $\theta'_i$ relative to
the beam direction by
\begin{equation}
Q_i^2 = -(p_i-p'_i)^2\approx 2E_i E'_i(1-\cos\theta'_i),
\label{eq-q2}
\end{equation}
where $p_i$ and $p'_i$ are the four-momenta of the beam electron and
the scattered electron, respectively, and $E_i$ and $E'_i$ are their
energies.  Events with detected scattered electrons (single-tagged or
double-tagged events) are excluded from the analysis. Driven by the
angular acceptance of the FD and SW calorimeters a value of
$Q^2=4.5$~GeV$^2$ is used in this analysis to separate the quasi-real
photons of untagged events from the virtual photons of tagged events.
The median $Q^2$ resulting from this definition cannot be determined
from data since the scattered electrons are not tagged. For the
kinematic range of this analysis the MC simulations predict the median
$Q^2$ to be of the order of $10^{-4}$~GeV$^2$.

The hadronic invariant mass of the photon-photon system, $W$, can be
obtained from the energies and momenta $(E_{\rm h} , \vec{p}_{\rm h})$
of final state particles:
\begin{eqnarray}
W^2 = (\sum_{\rm h}E_{\rm h})^2-(\sum_{{\rm h}}\vec{p}_{\rm h})^2.
\end{eqnarray}

The MC generators PYTHIA 5.722~\cite{bib-pythia} and PHOJET
1.10~\cite{bib-phojet} have been used to simulate photon-photon
interactions.  PYTHIA uses the SaS-1D parametrisation~\cite{bib-sas1d}
for the photon parton densities and PHOJET uses the GRV
parametrisation~\cite{bib-grv}.

All relevant background processes were studied using MC
generators. Multihadronic events ($\ee\rightarrow \qqbar(\gamma)$)
were simulated with PYTHIA 6.125.  KORALZ 4.02~\cite{bib-koralz} was
used to generate the process $\ee\rightarrow\tau^+\tau^-(\gamma)$ and
VERMASEREN~\cite{bib-vermaseren} to generate
$\ee\rightarrow\ee\tau^+\tau^-$.  Deep-inelastic e$\gamma$ events were
simulated with HERWIG 5.9~\cite{bib-herwig}.

All signal and background MC samples were generated with full
simulation of the OPAL detector~\cite{bib-gopal}.  They are analysed
using the same reconstruction algorithms as for the data.


\section{Event selection}
\label{sec-evsel}
The production of charged hadrons was studied using data taken at
$\ee$ centre-of-mass energies, $\sqee$, from 183 to 209~GeV, amounting
to a total integrated luminosity of 612.8~pb$^{-1}$.  The luminosity
weighted average centre-of-mass energy is 195.8~GeV.  Two-photon events are
selected with the following set of cuts:
\begin{itemize}

\item At least six tracks must have been found in the tracking chambers.
  A track is required to have a minimum transverse momentum of 120 MeV
  with respect to the $z$ axis and at least 40 hits in the central jet
  chamber. The number of measured hits in the jet chamber must be more
  than half of the number of possible hits given the track direction.
  The radial distance of nearest approach of the track to the primary
  vertex has to be less than 0.15~cm.

\item The visible invariant hadronic mass calculated from the position
  and the energy of the clusters measured in the ECAL has to be
  greater than 3~GeV.

\item The sum of all energy deposits in the ECAL and the HCAL has to
  be less than 50~GeV to remove background from hadronic Z decays in
  events with a radiative return to the Z peak.

\item The missing transverse momentum of the event measured in the
  ECAL and the FD has to be less than 8 GeV.

\item To reject events with scattered electrons in the FD or SW, the
  total energy sum measured in the FD and SW has to be less than
  60~GeV.

\item The background due to beam-gas or beam-wall interactions is
  reduced by requiring the radial distance of the primary vertex from
  the beam axis to be less than 2~cm and the distance from the nominal
  vertex position along the $z$ direction to be less than 3~cm.

\end{itemize}

After all cuts the data sample contains 1\,144\,035 events. The main
remaining background processes are multihadronic Z decays,
$\ee\longrightarrow \ee\tau^+\tau^-$ and deep-inelastic e$\gamma$
scattering. Other background processes are found to be negligible.
The multihadronic background is mainly reduced by the cut on the sum
of the energy measured by the ECAL and the HCAL and by the cut on the
missing transverse momentum.  The $\ee\longrightarrow \ee\tau^+\tau^-$
processes are reduced by the cuts on the number of tracks and
deep-inelastic e$\gamma$ events are rejected by the cut on the energy
in SW and FD. The trigger efficiency for events
in this region of phase space is close to 100\%~\cite{bib-opalhadr}
and no correction is applied.

From the MC simulations it is estimated that
after all cuts the total remaining background is below 2\% overall,
but increasing to up to 50\% at very high transverse momenta of the
charged particles. The background is subtracted bin-by-bin from the
distributions measured before corrections for detector resolution and
acceptance are applied. The signal MC generators PHOJET and PYTHIA are
found to underestimate the cross-section. In particular they show a
shape different from that in the data of the $\pt$-distribution of
charged hadrons. To study a potential bias resulting from this
deficiency the MC events have been reweighted by a suitable
$\pt$-dependent function to resemble the data. Both the original and
the reweighted MC distributions are used in the analysis and any
differences seen are included in the systematic uncertainty.


\section{Analysis}

Only particles with a lifetime $\tau>0.3$~ns are used to define the
primary charged hadronic multiplicity.  The primary charged hadrons
originate either directly from the primary interaction or from the
decay of particles with a lifetime $\tau<0.3$~ns. The decay products
of $\Lambda$ and $\ks$ particles are hence considered as primary
hadrons in this analysis.  The measured transverse momentum and
pseudorapidity distributions of charged hadrons have to be corrected
for the losses due to event and track selection cuts, the acceptance
and the resolution of the detector.  This is done using the MC events
which were processed by the full detector simulation and
reconstruction chain.  The data are corrected by multiplying the
experimental distribution, e.g.~of the transverse momentum $\pT$, with
correction factors which are calculated as the bin-by-bin ratio of the
generated and the reconstructed MC distributions:
\begin{equation}
\left(\frac{{\rm d}\sigma}{{\rm d} \pT}
\right)_{\rm corrected}=
\frac{\left(\frac{{\rm d}\sigma}{{\rm d} \pT}
\right)^{\rm MC}_{\rm generated~~~~}}{
\left(\frac{{\rm d}\sigma}{{\rm d} \pT}
\right)^{\rm MC}_{\rm reconstructed}}
\left(\frac{{\rm d}\sigma}{{\rm d} \pT}
\right)_{\rm measured}.
\label{eq1}
\end{equation}
The ratio is calculated using both PYTHIA and PHOJET, with the mean
value used to correct the data. The pseudorapidity distributions are
corrected in the same way.  This method only yields reliable results
if migration effects between bins due to the finite resolution of the
measurement are small. The bins of the $\pt$ and $|\eta|$
distributions have therefore been chosen to be significantly larger
than the resolution expected from the MC simulation.  To avoid regions
where the detector has little or no acceptance, all measurements of
charged hadron distributions were restricted to the range
$|\eta|<1.5$. The average transverse momentum $\langle \pt \rangle$
and the average pseudorapidity $\langle |\eta| \rangle$ in each bin is
calculated directly from the data since detector corrections are small
compared to the statistical uncertainties.

In order to be able to use Eq.~\ref{eq1} for the detector
correction in each bin of $W$, the bins in $W$ must be larger than
the experimental resolution and the average reconstructed hadronic
invariant mass, $\langle W_{\rm rec}\rangle$, should be
approximately equal to the average generated hadronic invariant
mass, $\langle W_{\rm gen}\rangle$.  In this case bin-to-bin
migrations are minimised. The visible invariant
mass, $W_{\rm vis}$, is determined from all tracks and calorimeter
clusters, including the forward detectors and the silicon tungsten
calorimeters, after applying a matching algorithm to avoid double 
counting of particle momenta~\cite{bib-mt}.

\begin{figure}[htpb]
   
\begin{center}
\epsfig{file=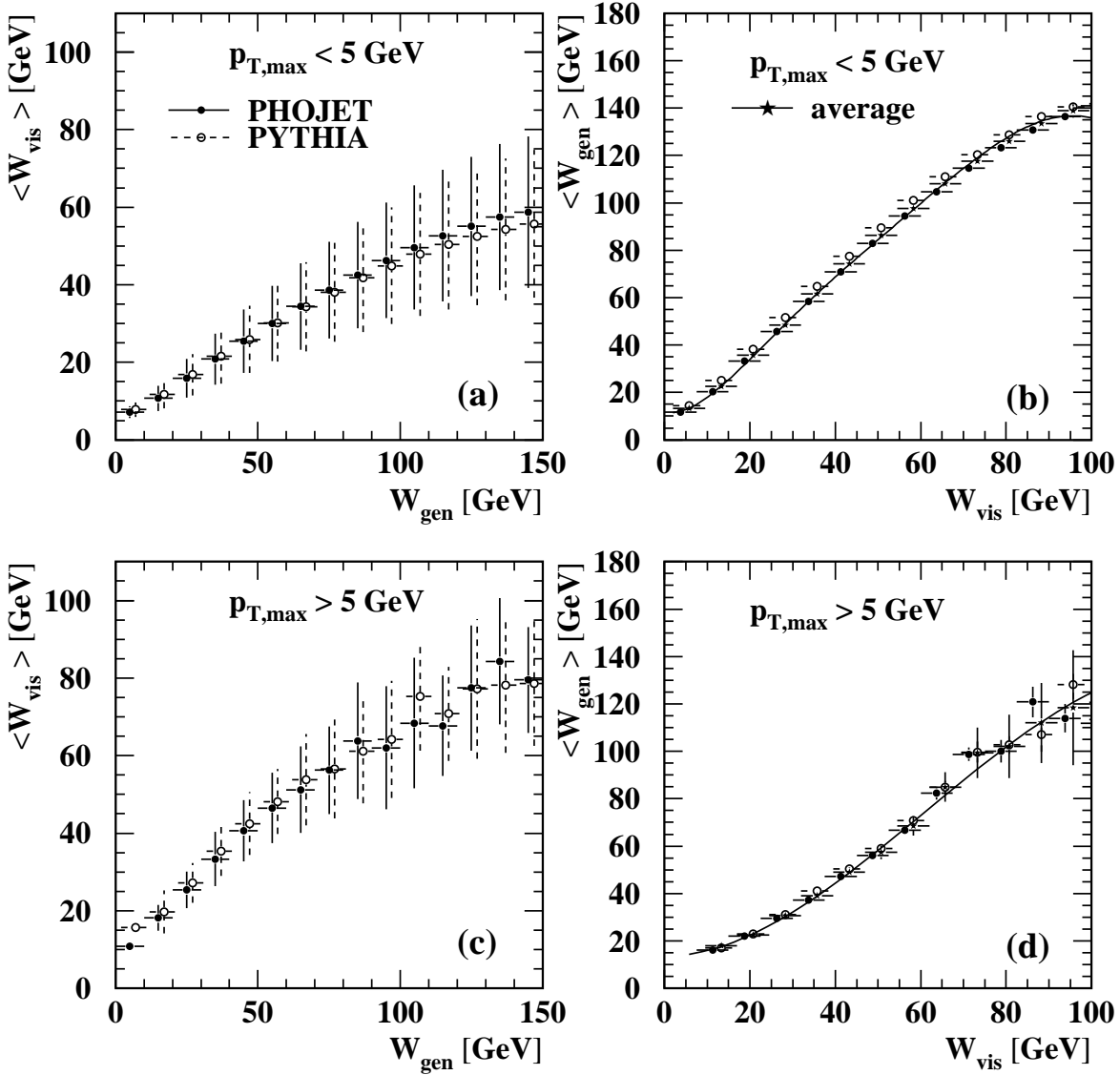,width=1.0\textwidth}
\end{center}
\caption{ The correlation between the generated hadronic invariant
mass $W_{\rm gen}$ and the visible hadronic invariant mass $W_{\rm
vis}$ in two regions of $p_{\rm T,max}$, the maximum transverse
momentum of any track in the event, for PHOJET and PYTHIA MC events.
The vertical bars show the standard deviation in each bin in (a,c) and
the uncertainty on the mean in (b,d) where larger than the marker
size.  The horizontal lines indicate the bin width. The polynomial fit
determines the correction function for $W_{\rm
vis}$. \label{fig-wcorr}}
\end{figure}

The ratio of $W_{\rm gen}$ and $W_{\rm vis}$ depends on the event
kinematics, and was therefore determined separately in two distinct
regions of phase space. The discriminating variable chosen for this
purpose is $p_{\rm T,max}$, the highest transverse momentum of any
track in the event.  The average visible hadronic invariant mass,
$\langle W_{\rm vis}\rangle$, and the resolution on $W_{\rm vis}$ as a
function of the generated hadronic invariant mass $W_{\rm gen}$ for
events with $p_{\rm T,max}$ smaller and larger than 5~GeV are shown in
Fig.~\ref{fig-wcorr}(a) and (c). The polynomial fits superimposed on
Fig.~\ref{fig-wcorr}(b) and (d) are used as a correction function so
that $\langle W_{\rm gen}\rangle/\langle W_{\rm rec}\rangle \approx
1$.

The distributions are measured for invariant masses $10<W<30$~GeV,
$30<W<50$~GeV, $50<W<125$~GeV and $10<W<125$~GeV where $W$ represents
the hadronic invariant mass after correcting for detector effects. To
facilitate comparisons to the results in~\cite{bib-l3}, the ranges
$W>30$~GeV and $W>50$~GeV are also considered.


\section{Systematic uncertainties}

The following sources of systematic uncertainties have been investigated:
\begin{itemize}

\item Significant amounts of background need to be subtracted from the
  data at large $\pt$, where the dominant source of background is
  multihadronic events ($\ee\rightarrow \qqbar(\gamma)$). The amount
  of this background subtracted from the data is varied by $\pm10$\%
  to estimate the uncertainty associated to this procedure. The amount
  by which the background must be varied is estimated from studying
  the MC description of the data in regions of the phase space where
  the background dominates. The effect on the measured cross-section
  is usually 1\,\% or less for low transverse momenta, increasing to
  up to 7\,\% at the highest momenta measured.

\item The selection criteria described in Section~\ref{sec-evsel} are
  varied simultaneously both to be more restrictive and to allow more
  events into the analysis to exclude a strong dependence on the event
  selection. Selection criteria based on energy measurements are
  varied by 5\% in the ECAL and HCAL, and by 10\% in the FD and SW
  calorimeters. The number of tracks required is changed by
  $\pm1$. The allowed radial distance of the tracks is varied by
  5\%. The uncertainty on the cross-section derived from all these
  variations is typically 1-6\%.

\item Both PHOJET and PYTHIA have been used to calculate the
  correction factors applied to the data. The resulting distributions
  are averaged and the difference between the two distributions is
  used to define the systematic uncertainty.  The distributions
  obtained from both programs have been reweighted for a better
  description of the data. The difference between using the reweighted
  and the original MC distributions to calculate the correction
  factors has been included in the systematic uncertainty.  The
  uncertainty derived from this study is below 5\%.

\item The systematic uncertainty due to the energy scale of the ECAL
  for the range of energies in this analysis was estimated by varying
  the reconstructed ECAL energy in the MC by
  $\pm3$\,\%~\cite{bib-totxs}.  The cross-sections change by up to 4\%
  due to this variation.

\item Studying vertex and net charge distributions it is estimated
  that about 2\% of the selected events are due to beam-gas or
  beam-wall interactions. This remaining background is treated as a
  systematic uncertainty of 2\%.

\end{itemize}

Systematic uncertainties due to the modelling of the detector
resolution for the measurement of tracks~\cite{bib-opalhadr} and due
to the luminosity measurement were found to be negligible.  The total
systematic uncertainty was obtained by adding all systematic
uncertainties in quadrature.


\section{Results and Conclusions}

\begin{figure}[htpb]
\begin{center}
\epsfig{file=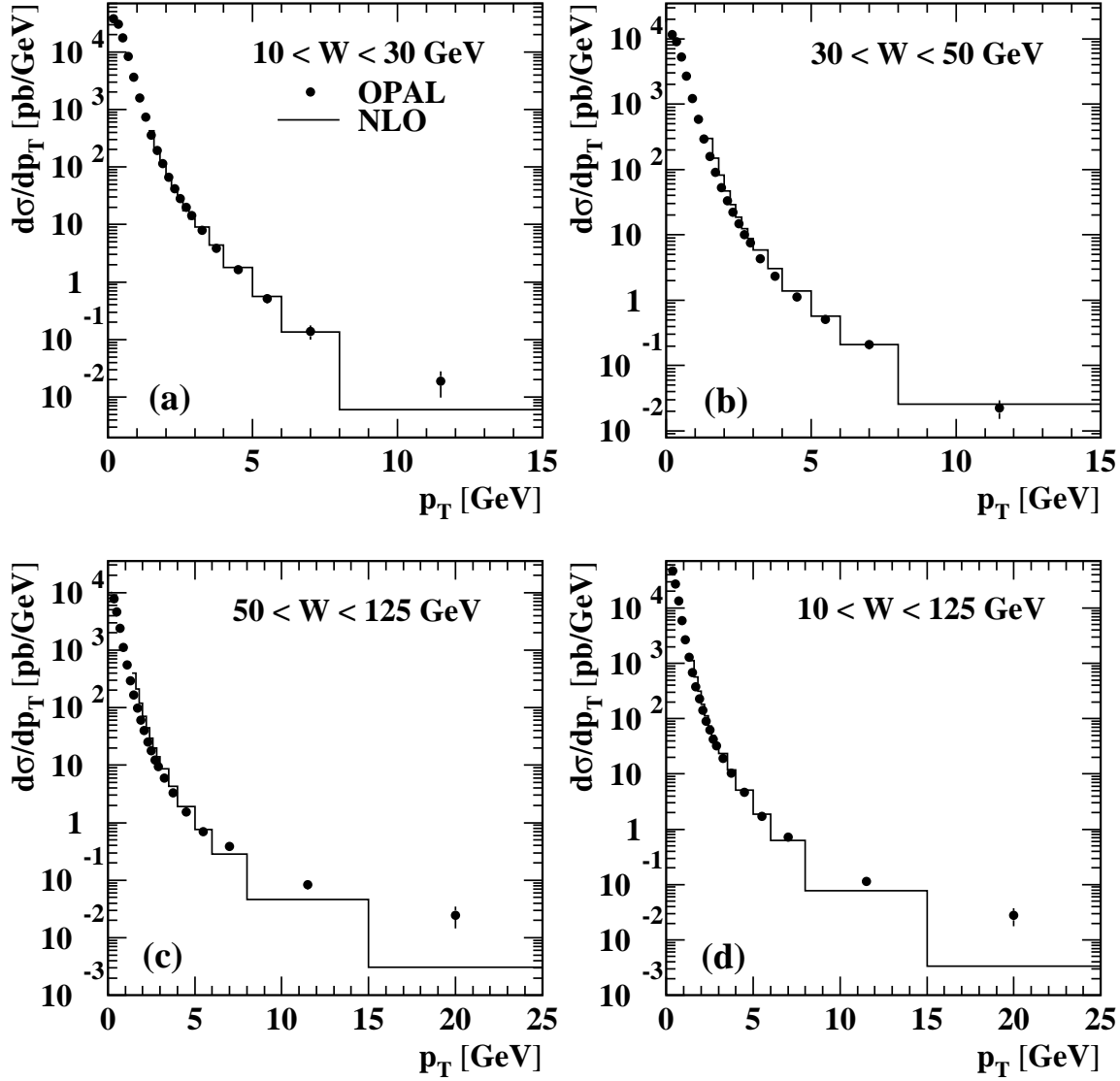,width=1.0\textwidth}
\end{center}
\caption{Differential inclusive charged hadron production cross-sections,
$\dspt$, for $|\eta|<1.5$ in the $W$ ranges 
(a) $10<W<30$~GeV;
(b) $30<W<50$~GeV;
(c) $50<~W<125$~GeV and 
(d) $10<W<125$~GeV. The error bars show
the statistical and systematic uncertainties added in quadrature when
larger than the marker, which indicates the bin centre. The data are compared to an NLO calculation~\cite{bib-binnewies}.
\label{fig-pt1}}
\end{figure}

The differential inclusive cross-sections ${\dspt}$ for charged
hadrons in four ranges of the hadronic invariant mass $W$ are shown in
Fig.~\ref{fig-pt1} and are given in Table~{\ref{tab-pt1a}} and
Table~{\ref{tab-pt1b}}. The data points in the figure indicate the bin
centre. Both data and calculations are presented for quasi-real
photons of virtualities $Q^2<4.5$~GeV$^2$, as described in
Sect.~\ref{sect-kin}.

Calculations in NLO QCD ~\cite{bib-binnewies} are compared to the
data.  The cross-sections are calculated using the QCD partonic
cross-sections in NLO for direct, single- and double-resolved
processes.  The hadronic cross-section is a convolution of the
Weizs\"acker-Williams effective photon distribution, the parton
distribution functions and the fragmentation functions
of~\cite{bib-ffs}. The AFG-HO parametrisation of the parton densities
of the photon~\cite{bib-pdf} is used with
$\Lambda^{5}_{\overline{MS}}= 221$~MeV. The renormalisation and
factorisation scales in the calculation are set equal to $p_{\rm T}$.
The cross-section calculation was repeated for the kinematic
conditions of the present analysis.  For the differential
cross-section $\dspt$ a minimum $p_{\rm T}$ of 1.5~GeV is required to
ensure the validity of the perturbative QCD calculation. Even at
$p_{\rm T}=1.5$~GeV the cross-sections change by up to 80\% when
varying the renormalisation and factorisation scales by factors of
two. This uncertainty decreases rapidly to between 10\% and 15\% for
$p_{\rm T}=3.5$~GeV and above. The differential cross-section $\dseta$
is hence restricted to the region $\pt>3.5$~GeV to allow a meaningful
comparison.

In Fig.~\ref{fig-pt1} the NLO calculation lies significantly below the
data for transverse momenta greater than about 10~GeV, which can be
reached in the highest $W$ range only. The predictions of PHOJET and
PYTHIA at high $\pt$ (not shown) are similar to the NLO calculation.

\begin{figure}[htbp]
\begin{center}
\vspace{5.mm}
\epsfig{file=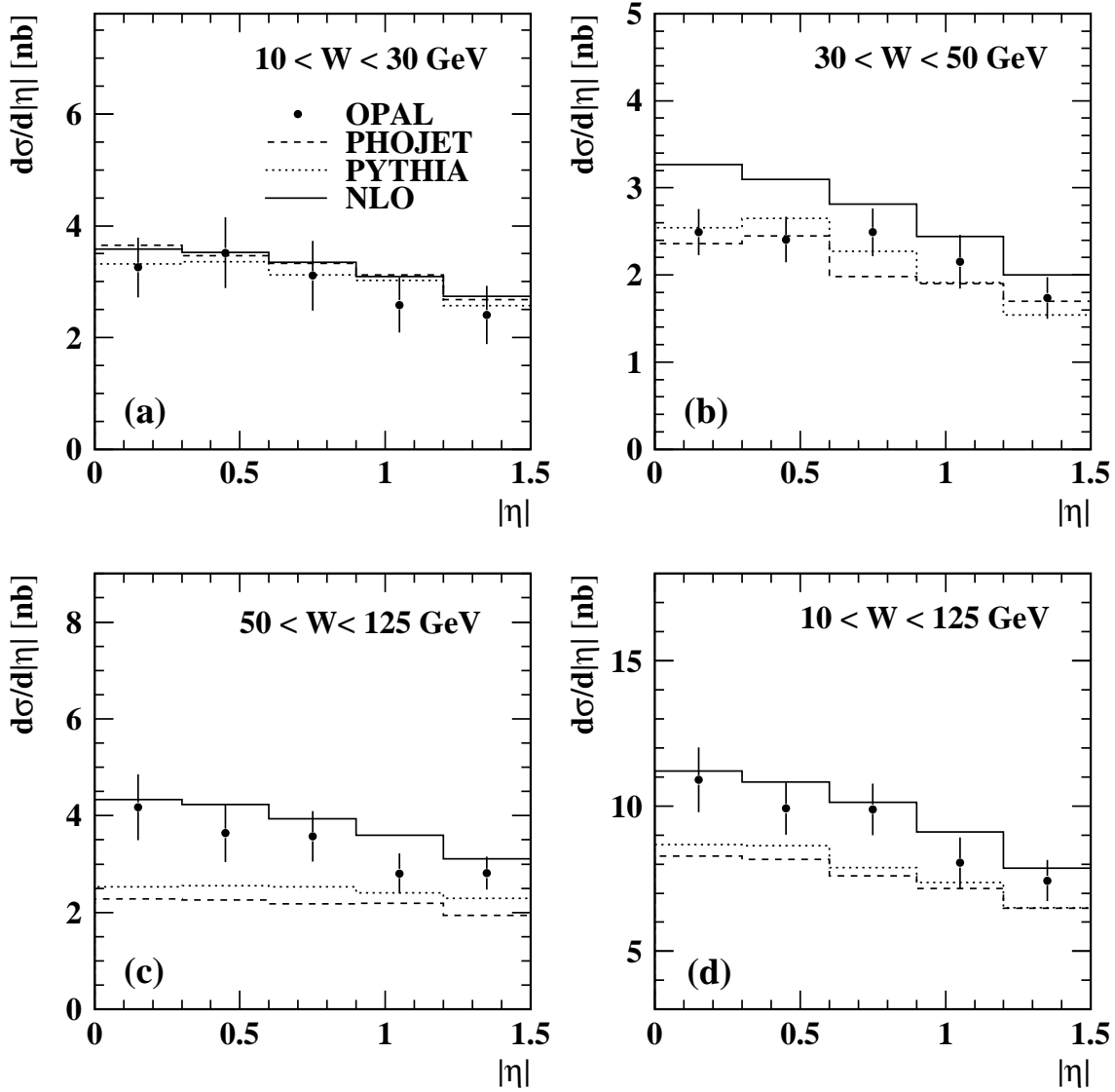,width=1.0\textwidth}
\end{center}
 \caption{Differential inclusive charged hadron production
cross-sections, $\dseta$, for $\pt>~3.5$~GeV and in the $W$ ranges
(a) $10<W<30$~GeV; (b) $30<W<50$~GeV; (c) $50<W<~125$~GeV and (d)
$10<W<125$~GeV.  The
data are compared to the PHOJET and PYTHIA models. 
The data are also compared to an NLO calculation~\cite{bib-binnewies}. 
The error bars show the statistical and systematic uncertainties added in
quadrature.The markers are shown at the centre of the bin.\label{fig-eta2}}
\end{figure}

The differential cross-section is nearly independent of $|\eta|$ in
the range measured as can be seen in Fig.~\ref{fig-eta2}. The data
points in the figure indicate the bin centre. The cross-section values
are given in Table~{\ref{tab-eta1}}. The NLO calculation reproduces
the data well within the uncertainties of the calculation. PYTHIA and
PHOJET describe the shape of the distributions correctly for all
regions of the phase space measured, but are below the data in
normalisation for large values of $W$.

\begin{figure}[htpb]
\begin{center}
\epsfig{file=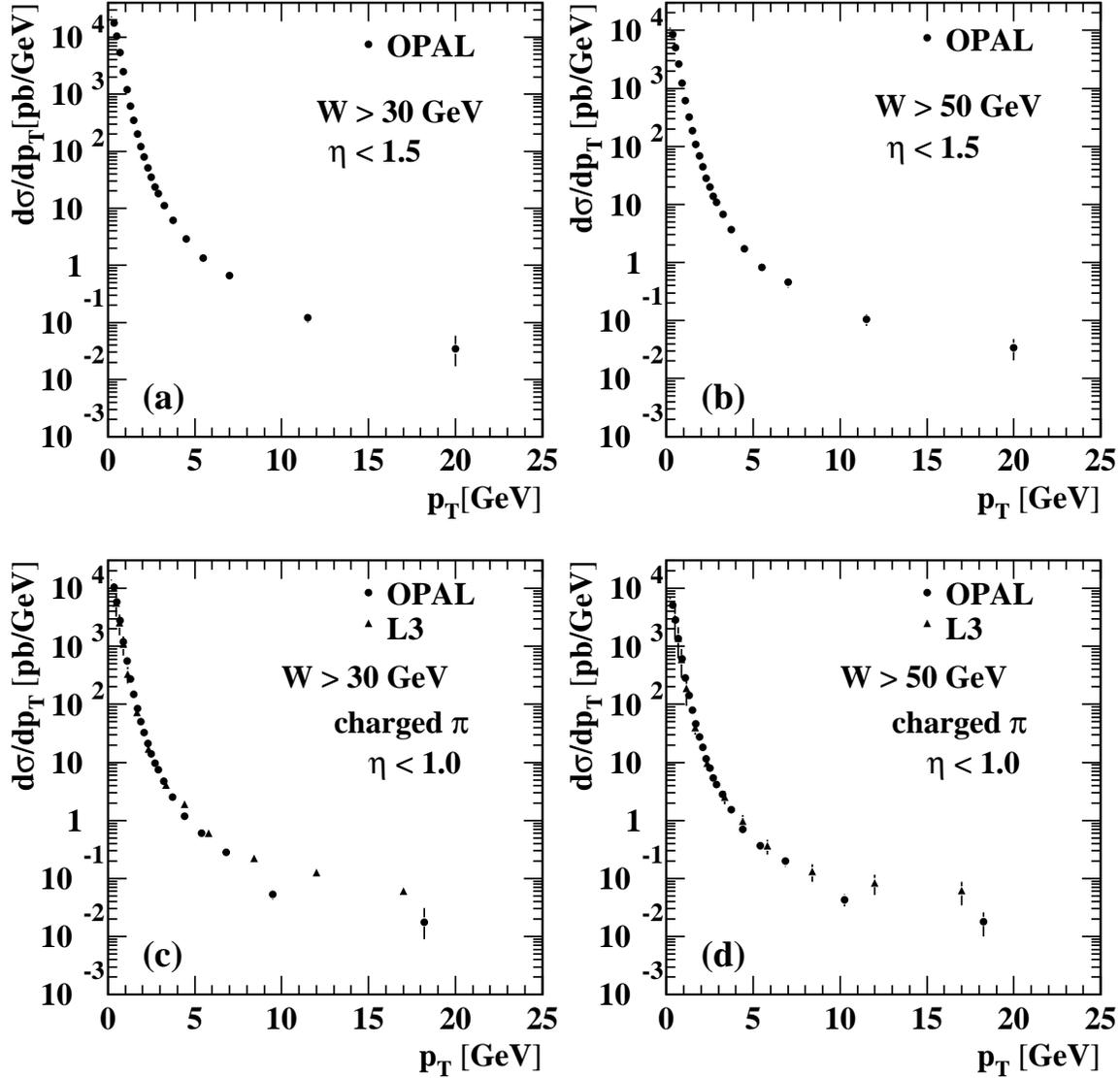,width=1.0\textwidth}
\end{center}
\caption{Differential inclusive charged hadron production
cross-sections, $\dspt$, for $|\eta|<1.5$ and in the $W$ ranges (a)
$W>30$~GeV and (b) $W>50$~GeV. 
The differential inclusive charged pion production cross-sections
$\dspt$ for $|\eta|<1.0$ in the $W$ ranges (c) $W>30$~GeV and (d)
$W>50$~GeV are inferred from the charged hadron cross-sections using
MC information to facilitate the comparison to the L3
measurements~\cite{bib-l3}. The error bars show the statistical and
systematic uncertainties added in quadrature when larger than the
marker. While the OPAL data points in (a) and (b) are plotted at the
bin centre, in (c) and (d) they are plotted at the mean transverse
momentum across the bin, as for the L3 data. \label{fig-pt2}}
\end{figure}

Fig.~\ref{fig-pt2} (a) and (b) and Table \ref{tab-ptl3} show the
differential cross-section ${\dspt}$ for charged hadrons for
$W>30$~GeV and $W>50$~GeV to facilitate a comparison with a recent
measurement by L3 of charged pions in the pseudorapidity range
$|\eta|<1.0$~\cite{bib-l3}.  The OPAL data shown in Fig.~\ref{fig-pt2}
(c) and (d) and Table \ref{tab-ptl3b} have been scaled to account for
the reduced $|\eta|$ range and for the fraction of charged pions of
all charged hadrons using MC simulations.  While the OPAL data points
in Fig.~\ref{fig-pt2} (a) and (b) are plotted at the bin centre, in
Fig.~\ref{fig-pt2} (c) and (d) they are plotted at the mean transverse
momentum across the bin, as for the L3 data.  From this comparison it
is evident that the distributions measured by OPAL fall more rapidly
towards high transverse momenta than those measured by L3, leading to
a disagreement between the two experiments at high transverse momenta
and a better description of the OPAL data by NLO QCD than is the case
for the L3 data.

\vspace{-4mm}
\section*{Acknowledgements}
We thank Bernd Kniehl for providing the NLO calculations and for 
many useful discussions. We particularly wish to thank the SL Division
for the efficient operation 
of the LEP accelerator at all energies
 and for their close cooperation with
our experimental group.  In addition to the support staff at our own
institutions we are pleased to acknowledge the  \\
Department of Energy, USA, \\
National Science Foundation, USA, \\
Particle Physics and Astronomy Research Council, UK, \\
Natural Sciences and Engineering Research Council, Canada, \\
Israel Science Foundation, administered by the Israel
Academy of Science and Humanities, \\
Benoziyo Center for High Energy Physics,\\
Japanese Ministry of Education, Culture, Sports, Science and
Technology (MEXT) and a grant under the MEXT International
Science Research Program,\\
Japanese Society for the Promotion of Science (JSPS),\\
German Israeli Bi-national Science Foundation (GIF), \\
Bundesministerium f\"ur Bildung und Forschung, Germany, \\
National Research Council of Canada, \\
Hungarian Foundation for Scientific Research, OTKA T-038240, 
and T-042864,\\
The NWO/NATO Fund for Scientific Research, the Netherlands.\\

%

\begin{table}
\begin{center} 
\begin{tabular}{|c||l|l||l|l|} \hline
 & \multicolumn{2}{c||}{$10<W<30$~GeV} & \multicolumn{2}{c|}{$30<W<50$~GeV} 
 \\[0.1cm]  \hline
$\pt$ & $\langle \pt \rangle$ & 
\multicolumn{1}{c||}{$\dspt$} & $\langle \pt \rangle$&
\multicolumn{1}{c|}{$\dspt$} \\
{[GeV]} & {[GeV]} & 
\multicolumn{1}{c||}{[pb/GeV]} & {[GeV]} &
\multicolumn{1}{c|}{[pb/GeV]} \\ \hline
 0.12--0.28 &  0.20    & (3.78\,$\pm$\,0.01\,$\pm$\,0.18)$\times 10^{ 4}$ &  0.20   & (1.15\,$\pm$\,0.00\,$\pm$\,0.03)$\times 10^{ 4}$\\
 0.28--0.44 &  0.35    & (3.01\,$\pm$\,0.01\,$\pm$\,0.13)$\times 10^{ 4}$ &  0.35   & (9.01\,$\pm$\,0.02\,$\pm$\,0.21)$\times 10^{ 3}$\\
 0.44--0.60 &  0.51    & (1.74\,$\pm$\,0.00\,$\pm$\,0.06)$\times 10^{ 4}$ &  0.51   & (5.26\,$\pm$\,0.02\,$\pm$\,0.13)$\times 10^{ 3}$\\
 0.60--0.80 &  0.69    & (8.43\,$\pm$\,0.03\,$\pm$\,0.29)$\times 10^{ 3}$ &  0.69   & (2.69\,$\pm$\,0.01\,$\pm$\,0.07)$\times 10^{ 3}$\\
 0.80--1.00 &  0.89    & (3.62\,$\pm$\,0.02\,$\pm$\,0.16)$\times 10^{ 3}$ &  0.89   & (1.22\,$\pm$\,0.01\,$\pm$\,0.03)$\times 10^{ 3}$\\
 1.00--1.20 &  1.09    & (1.58\,$\pm$\,0.01\,$\pm$\,0.09)$\times 10^{ 3}$ &  1.09   & (5.85\,$\pm$\,0.05\,$\pm$\,0.19)$\times 10^{ 2}$\\
 1.20--1.40 &  1.29    & (7.29\,$\pm$\,0.08\,$\pm$\,0.48)$\times 10^{ 2}$ &  1.29   & (2.94\,$\pm$\,0.04\,$\pm$\,0.11)$\times 10^{ 2}$\\
 1.40--1.60 &  1.49    & (3.61\,$\pm$\,0.06\,$\pm$\,0.25)$\times 10^{ 2}$ &  1.49   & (1.59\,$\pm$\,0.03\,$\pm$\,0.06)$\times 10^{ 2}$\\
 1.60--1.80 &  1.69    & (1.92\,$\pm$\,0.04\,$\pm$\,0.13)$\times 10^{ 2}$ &  1.69   & (9.13\,$\pm$\,0.21\,$\pm$\,0.32)$\times 10^{ 1}$\\
 1.80--2.00 &  1.89    & (1.13\,$\pm$\,0.03\,$\pm$\,0.08)$\times 10^{ 2}$ &  1.89   & (5.32\,$\pm$\,0.15\,$\pm$\,0.19)$\times 10^{ 1}$\\
 2.00--2.20 &  2.09    & (6.68\,$\pm$\,0.23\,$\pm$\,0.44)$\times 10^{ 1}$ &  2.09   & (3.34\,$\pm$\,0.12\,$\pm$\,0.11)$\times 10^{ 1}$\\
 2.20--2.40 &  2.29    & (4.17\,$\pm$\,0.19\,$\pm$\,0.26)$\times 10^{ 1}$ &  2.29   & (2.22\,$\pm$\,0.10\,$\pm$\,0.07)$\times 10^{ 1}$\\
 2.40--2.60 &  2.50    & (2.85\,$\pm$\,0.16\,$\pm$\,0.17)$\times 10^{ 1}$ &  2.50   & (1.47\,$\pm$\,0.08\,$\pm$\,0.04)$\times 10^{ 1}$\\
 2.60--2.80 &  2.70    & (1.98\,$\pm$\,0.13\,$\pm$\,0.12)$\times 10^{ 1}$ &  2.70   & (1.00\,$\pm$\,0.06\,$\pm$\,0.03)$\times 10^{ 1}$\\
 2.80--3.00 &  2.90    & (1.45\,$\pm$\,0.12\,$\pm$\,0.09)$\times 10^{ 1}$ &  2.90   & (7.48\,$\pm$\,0.54\,$\pm$\,0.26)$\times 10^{ 0}$\\
 3.00--3.50 &  3.21    & (7.93\,$\pm$\,0.56\,$\pm$\,0.50)$\times 10^{ 0}$ &  3.23   & (4.33\,$\pm$\,0.26\,$\pm$\,0.15)$\times 10^{ 0}$\\
 3.50--4.00 &  3.71    & (3.87\,$\pm$\,0.42\,$\pm$\,0.27)$\times 10^{ 0}$ &  3.73   & (2.32\,$\pm$\,0.19\,$\pm$\,0.08)$\times 10^{ 0}$\\
 4.00--5.00 &  4.40    & (1.63\,$\pm$\,0.22\,$\pm$\,0.15)$\times 10^{ 0}$ &  4.42   & (1.12\,$\pm$\,0.09\,$\pm$\,0.04)$\times 10^{ 0}$\\
 5.00--6.00 &  5.40    & (5.16\,$\pm$\,0.78\,$\pm$\,0.55)$\times 10^{-1}$ &  5.43   & (5.05\,$\pm$\,0.66\,$\pm$\,0.19)$\times 10^{-1}$\\
 6.00--8.00 &  6.74\,$\pm$\,0.01    & (1.37\,$\pm$\,0.31\,$\pm$\,0.22)$\times 10^{-1}$ &  6.74   & (2.10\,$\pm$\,0.33\,$\pm$\,0.12)$\times 10^{-1}$\\
 8.00--15.00 &  9.52\,$\pm$\,0.12    & (1.90\,$\pm$\,0.86\,$\pm$\,0.33)$\times 10^{-2}$ &  9.62\,$\pm$\,0.03   & (2.23\,$\pm$\,0.60\,$\pm$\,0.36)$\times 10^{-2}$\\
\hline
\end{tabular}
\caption{Differential inclusive charged hadron production
  cross-sections $\dspt$ for $|\eta|<1.5$ and in the $W$ ranges
  $10<W<30$~GeV and $30<W<50$~GeV.  The first uncertainty is the
  statistical uncertainty and the second uncertainty is the systematic
  uncertainty. No value is given if the error on $\langle \pt
  \rangle$ is less than 0.01. }
\label{tab-pt1a}
\end{center}
\end{table}

\begin{table}[htpb]
\begin{center} 
\begin{tabular}{|c||l|l||l|l|} \hline
 & \multicolumn{2}{c||}{$50<W<125$~GeV} & \multicolumn{2}{c|}{$10<W<125$~GeV} 
 \\[0.1cm]  \hline
$\pt$ & $\langle \pt \rangle$ & 
\multicolumn{1}{c||}{$\dspt$} & $\langle \pt \rangle$ &
\multicolumn{1}{c|}{$\dspt$~[pb/GeV]} \\ 
{[GeV]} & {[GeV]} & 
\multicolumn{1}{c||}{[pb/GeV]} & [GeV] &
\multicolumn{1}{c|}{[pb/GeV]} \\ \hline
 0.12--0.28 &  0.20    & (1.00\,$\pm$\,0.00\,$\pm$\,0.03)$\times 10^{ 4}$ &  0.20   & (5.93\,$\pm$\,0.00\,$\pm$\,0.21)$\times 10^{ 4}$\\
 0.28--0.44 &  0.35    & (7.77\,$\pm$\,0.02\,$\pm$\,0.22)$\times 10^{ 3}$ &  0.35   & (4.68\,$\pm$\,0.00\,$\pm$\,0.18)$\times 10^{ 4}$\\
 0.44--0.60 &  0.51    & (4.58\,$\pm$\,0.01\,$\pm$\,0.13)$\times 10^{ 3}$ &  0.51   & (2.71\,$\pm$\,0.00\,$\pm$\,0.10)$\times 10^{ 4}$\\
 0.60--0.80 &  0.69    & (2.37\,$\pm$\,0.01\,$\pm$\,0.07)$\times 10^{ 3}$ &  0.69   & (1.34\,$\pm$\,0.00\,$\pm$\,0.05)$\times 10^{ 4}$\\
 0.80--1.00 &  0.89    & (1.12\,$\pm$\,0.01\,$\pm$\,0.03)$\times 10^{ 3}$ &  0.89   & (5.87\,$\pm$\,0.01\,$\pm$\,0.25)$\times 10^{ 3}$\\
 1.00--1.20 &  1.09    & (5.56\,$\pm$\,0.04\,$\pm$\,0.14)$\times 10^{ 2}$ &  1.09   & (2.68\,$\pm$\,0.01\,$\pm$\,0.14)$\times 10^{ 3}$\\
 1.20--1.40 &  1.29    & (2.92\,$\pm$\,0.03\,$\pm$\,0.08)$\times 10^{ 2}$ &  1.29   & (1.29\,$\pm$\,0.01\,$\pm$\,0.07)$\times 10^{ 3}$\\
 1.40--1.60 &  1.49    & (1.66\,$\pm$\,0.02\,$\pm$\,0.05)$\times 10^{ 2}$ &  1.49   & (6.78\,$\pm$\,0.04\,$\pm$\,0.38)$\times 10^{ 2}$\\
 1.60--1.80 &  1.69    & (9.74\,$\pm$\,0.15\,$\pm$\,0.30)$\times 10^{ 1}$ &  1.69   & (3.78\,$\pm$\,0.03\,$\pm$\,0.20)$\times 10^{ 2}$\\
 1.80--2.00 &  1.89    & (6.10\,$\pm$\,0.11\,$\pm$\,0.21)$\times 10^{ 1}$ &  1.89   & (2.27\,$\pm$\,0.02\,$\pm$\,0.12)$\times 10^{ 2}$\\
 2.00--2.20 &  2.09    & (4.01\,$\pm$\,0.09\,$\pm$\,0.15)$\times 10^{ 1}$ &  2.09   & (1.41\,$\pm$\,0.02\,$\pm$\,0.07)$\times 10^{ 2}$\\
 2.20--2.40 &  2.29    & (2.53\,$\pm$\,0.07\,$\pm$\,0.10)$\times 10^{ 1}$ &  2.29   & (9.03\,$\pm$\,0.16\,$\pm$\,0.44)$\times 10^{ 1}$\\
 2.40--2.60 &  2.50    & (1.76\,$\pm$\,0.06\,$\pm$\,0.09)$\times 10^{ 1}$ &  2.50   & (6.18\,$\pm$\,0.13\,$\pm$\,0.31)$\times 10^{ 1}$\\
 2.60--2.80 &  2.70    & (1.22\,$\pm$\,0.05\,$\pm$\,0.06)$\times 10^{ 1}$ &  2.70   & (4.26\,$\pm$\,0.11\,$\pm$\,0.22)$\times 10^{ 1}$\\
 2.80--3.00 &  2.90    & (9.52\,$\pm$\,0.44\,$\pm$\,0.50)$\times 10^{ 0}$ &  2.90   & (3.25\,$\pm$\,0.09\,$\pm$\,0.17)$\times 10^{ 1}$\\
 3.00--3.50 &  3.23    & (5.99\,$\pm$\,0.22\,$\pm$\,0.30)$\times 10^{ 0}$ &  3.23   & (1.93\,$\pm$\,0.05\,$\pm$\,0.10)$\times 10^{ 1}$\\
 3.50--4.00 &  3.73    & (3.33\,$\pm$\,0.16\,$\pm$\,0.22)$\times 10^{ 0}$ &  3.73   & (1.03\,$\pm$\,0.03\,$\pm$\,0.06)$\times 10^{ 1}$\\
 4.00--5.00 &  4.40    & (1.52\,$\pm$\,0.08\,$\pm$\,0.15)$\times 10^{ 0}$ &  4.43   & (4.66\,$\pm$\,0.16\,$\pm$\,0.34)$\times 10^{ 0}$\\
 5.00--6.00 &  5.43    & (7.02\,$\pm$\,0.83\,$\pm$\,0.81)$\times 10^{-1}$ &  5.43   & (1.71\,$\pm$\,0.11\,$\pm$\,0.13)$\times 10^{ 0}$\\
 6.00--8.00 &  6.83    & (3.89\,$\pm$\,0.45\,$\pm$\,0.49)$\times 10^{-1}$ &  6.79   & (7.28\,$\pm$\,0.53\,$\pm$\,0.62)$\times 10^{-1}$\\
 8.00--15.00 & 10.18\,$\pm$\,0.01    & (8.40\,$\pm$\,1.34\,$\pm$\,0.90)$\times 10^{-2}$ & 10.00   & (1.14\,$\pm$\,0.14\,$\pm$\,0.09)$\times 10^{-1}$\\
15.00--25.00 & 18.26\,$\pm$\,0.09    & (2.46\,$\pm$\,0.96\,$\pm$\,0.26)$\times 10^{-2}$ & 18.27\,$\pm$\,0.08   & (2.76\,$\pm$\,0.99\,$\pm$\,0.22)$\times 10^{-2}$\\
\hline
\end{tabular}
\caption{Differential inclusive charged hadron production
  cross-sections $\dspt$ for $|\eta|<1.5$ and in the $W$ ranges
  $50<W<125$~GeV and $10<W<125$~GeV.  The first uncertainty is the
  statistical uncertainty and the second uncertainty is the systematic
  uncertainty. No value is given if the error on $\langle \pt \rangle$
  is less than 0.01.}
\label{tab-pt1b}
\end{center}
\end{table}

\begin{table}[htpb]
\begin{center} 
\begin{tabular}{|c||l|l||l|l|} \hline
 & \multicolumn{2}{|c||}{$10<W<30$ GeV} & \multicolumn{2}{c|}{$30<W<50$ GeV} \\[0.1cm]  \hline
$|\eta|$ & $\langle |\eta| \rangle$ &\multicolumn{1}{c||}{$\dseta$~[pb]} &
$\langle  |\eta | \rangle$ &\multicolumn{1}{c|}{$\dseta$~[pb]} \\ \hline
 0.0--0.3 & 0.149 & 3.26$\pm$0.45$\pm$0.29 & 0.151 & 2.49$\pm$0.26$\pm$0.07 \\
 0.3--0.6 & 0.445 & 3.52$\pm$0.50$\pm$0.39 & 0.451 & 2.41$\pm$0.25$\pm$0.08 \\
 0.6--0.9 & 0.754 & 3.11$\pm$0.47$\pm$0.41 & 0.759 & 2.48$\pm$0.26$\pm$0.07 \\
 0.9--1.2 & 1.047 & 2.58$\pm$0.43$\pm$0.24 & 1.049 & 2.15$\pm$0.26$\pm$0.17 \\
 1.2--1.5 & 1.348 & 2.40$\pm$0.44$\pm$0.28 & 1.341 & 1.74$\pm$0.23$\pm$0.07 \\
 \hline
\multicolumn{5}{c}{}\\
\hline
  & \multicolumn{2}{|c||}{$50<W<125$ GeV} & \multicolumn{2}{c|}{$10<W<125$ GeV} \\[0.1cm]  \hline
$|\eta|$  & $\langle |\eta| \rangle$ & \multicolumn{1}{c||}{$\dseta$~[pb]} &
$\langle  |\eta | \rangle$ &\multicolumn{1}{c|}{$\dseta$~[pb]} \\ \hline
 0.0--0.3 & 0.149 & 4.17$\pm$0.27$\pm$0.63 & 0.149 & 10.90$\pm$0.65$\pm$0.90  \\
 0.3--0.6 & 0.449 & 3.63$\pm$0.25$\pm$0.53 & 0.448 & 9.92\ $\pm$0.62$\pm$0.65 \\
 0.6--0.9 & 0.759 & 3.57$\pm$0.25$\pm$0.46 & 0.757 & 9.89\ $\pm$0.63$\pm$0.63 \\
 0.9--1.2 & 1.048 & 2.81$\pm$0.23$\pm$0.34 & 1.047 & 8.04\ $\pm$0.58$\pm$0.67 \\
 1.2--1.5 & 1.345 & 2.81$\pm$0.22$\pm$0.25 & 1.345 & 7.42\ $\pm$0.55$\pm$0.44 \\
\hline
\end{tabular}
\caption{Differential inclusive charged hadron production cross-sections
$\dseta$ for $\pt~>~3.5$~GeV and in the $W$ ranges 
$10<W<30$~GeV, $30<W<50$~GeV,
$50<W<125$~GeV and $10<W<125$~GeV.
The first uncertainty is the statistical uncertainty and
the second uncertainty is the systematic uncertainty.}
\label{tab-eta1}
\end{center}
\end{table}

\begin{table}[htpb]
\begin{center} 
\begin{tabular}{|c||l|l||l|l|} \hline
 & \multicolumn{2}{c||}{$W>30$~GeV} & \multicolumn{2}{c|}{$W>50$~GeV} 
 \\[0.1cm]  \hline
$\pt$ & $\langle \pt \rangle$  & 
\multicolumn{1}{c||}{$\dspt$} & $\langle \pt \rangle$ &
\multicolumn{1}{c|}{$\dspt$} \\
{[GeV]} &  {[GeV]} & 
\multicolumn{1}{c||}{[pb/GeV]} & [GeV] &
\multicolumn{1}{c|}{[pb/GeV]} \\ \hline
 0.12--0.28 &  0.20    & (2.25\,$\pm$\,0.00\,$\pm$\,0.06)$\times 10^{ 4}$ &  0.20   & (1.10\,$\pm$\,0.00\,$\pm$\,0.04)$\times 10^{ 4}$\\
 0.28--0.44 &  0.35    & (1.75\,$\pm$\,0.00\,$\pm$\,0.05)$\times 10^{ 4}$ &  0.35   & (8.50\,$\pm$\,0.02\,$\pm$\,0.28)$\times 10^{ 3}$\\
 0.44--0.60 &  0.51    & (1.03\,$\pm$\,0.00\,$\pm$\,0.03)$\times 10^{ 4}$ &  0.51   & (5.02\,$\pm$\,0.02\,$\pm$\,0.16)$\times 10^{ 3}$\\
 0.60--0.80 &  0.69    & (5.31\,$\pm$\,0.02\,$\pm$\,0.13)$\times 10^{ 3}$ &  0.69   & (2.61\,$\pm$\,0.01\,$\pm$\,0.08)$\times 10^{ 3}$\\
 0.80--1.00 &  0.89    & (2.46\,$\pm$\,0.01\,$\pm$\,0.06)$\times 10^{ 3}$ &  0.89   & (1.24\,$\pm$\,0.01\,$\pm$\,0.04)$\times 10^{ 3}$\\
 1.00--1.20 &  1.09    & (1.20\,$\pm$\,0.01\,$\pm$\,0.03)$\times 10^{ 3}$ &  1.09   & (6.17\,$\pm$\,0.05\,$\pm$\,0.18)$\times 10^{ 2}$\\
 1.20--1.40 &  1.29    & (6.19\,$\pm$\,0.05\,$\pm$\,0.16)$\times 10^{ 2}$ &  1.29   & (3.25\,$\pm$\,0.04\,$\pm$\,0.09)$\times 10^{ 2}$\\
 1.40--1.60 &  1.49    & (3.44\,$\pm$\,0.04\,$\pm$\,0.09)$\times 10^{ 2}$ &  1.49   & (1.85\,$\pm$\,0.03\,$\pm$\,0.05)$\times 10^{ 2}$\\
 1.60--1.80 &  1.69    & (2.00\,$\pm$\,0.03\,$\pm$\,0.06)$\times 10^{ 2}$ &  1.69   & (1.09\,$\pm$\,0.02\,$\pm$\,0.03)$\times 10^{ 2}$\\
 1.80--2.00 &  1.89    & (1.22\,$\pm$\,0.02\,$\pm$\,0.04)$\times 10^{ 2}$ &  1.89   & (6.87\,$\pm$\,0.18\,$\pm$\,0.20)$\times 10^{ 1}$\\
 2.00--2.20 &  2.09    & (7.86\,$\pm$\,0.19\,$\pm$\,0.24)$\times 10^{ 1}$ &  2.09   & (4.51\,$\pm$\,0.14\,$\pm$\,0.14)$\times 10^{ 1}$\\
 2.20--2.40 &  2.29    & (5.09\,$\pm$\,0.15\,$\pm$\,0.16)$\times 10^{ 1}$ &  2.29   & (2.86\,$\pm$\,0.11\,$\pm$\,0.10)$\times 10^{ 1}$\\
 2.40--2.60 &  2.50    & (3.48\,$\pm$\,0.12\,$\pm$\,0.13)$\times 10^{ 1}$ &  2.50   & (2.01\,$\pm$\,0.09\,$\pm$\,0.08)$\times 10^{ 1}$\\
 2.60--2.80 &  2.70    & (2.38\,$\pm$\,0.10\,$\pm$\,0.09)$\times 10^{ 1}$ &  2.70   & (1.38\,$\pm$\,0.08\,$\pm$\,0.06)$\times 10^{ 1}$\\
 2.80--3.00 &  2.90    & (1.83\,$\pm$\,0.09\,$\pm$\,0.07)$\times 10^{ 1}$ &  2.90   & (1.08\,$\pm$\,0.07\,$\pm$\,0.05)$\times 10^{ 1}$\\
 3.00--3.50 &  3.23    & (1.12\,$\pm$\,0.04\,$\pm$\,0.04)$\times 10^{ 1}$ &  3.23   & (6.78\,$\pm$\,0.34\,$\pm$\,0.27)$\times 10^{ 0}$\\
 3.50--4.00 &  3.73    & (6.08\,$\pm$\,0.32\,$\pm$\,0.25)$\times 10^{ 0}$ &  3.73   & (3.72\,$\pm$\,0.26\,$\pm$\,0.17)$\times 10^{ 0}$\\
 4.00--5.00 &  4.43    & (2.89\,$\pm$\,0.16\,$\pm$\,0.16)$\times 10^{ 0}$ &  4.43   & (1.74\,$\pm$\,0.12\,$\pm$\,0.13)$\times 10^{ 0}$\\
 5.00--6.00 &  5.43    & (1.35\,$\pm$\,0.14\,$\pm$\,0.08)$\times 10^{ 0}$ &  5.43   & (8.23\,$\pm$\,1.22\,$\pm$\,0.73)$\times 10^{-1}$\\
 6.00--8.00 &  6.80    & (6.61\,$\pm$\,0.79\,$\pm$\,0.48)$\times 10^{-1}$ &  6.83   & (4.57\,$\pm$\,0.72\,$\pm$\,0.44)$\times 10^{-1}$\\
 8.00--15.00 & 10.11    & (1.22\,$\pm$\,0.20\,$\pm$\,0.09)$\times 10^{-1}$ & 10.25   & (1.04\,$\pm$\,0.21\,$\pm$\,0.10)$\times 10^{-1}$\\
15.00--25.00 & 18.16\,$\pm$\,0.03    & (3.43\,$\pm$\,2.46\,$\pm$\,0.27)$\times 10^{-2}$ & 18.14\,$\pm$\,0.03   & (3.43\,$\pm$\,1.34\,$\pm$\,0.32)$\times 10^{-2}$\\
\hline
\end{tabular}
\caption{Differential inclusive charged hadron production
  cross-sections $\dspt$ for $|\eta|<1.5$ and in the ranges $W>30$~GeV
  and $W>50$~GeV.  The first uncertainty is the statistical
  uncertainty and the second uncertainty is the systematic
  uncertainty. No value is given if the error on $\langle \pt \rangle$
  is less than 0.01.}
\label{tab-ptl3}
\end{center}
\end{table}

\newpage
\begin{table}[htpb]
\begin{center} 
\begin{tabular}{|c||l|l||l|l|} \hline
 & \multicolumn{2}{c||}{$W>30$~GeV} & \multicolumn{2}{c|}{$W>50$~GeV}\\[0.1cm] 
\hline
$\pt$  & $\langle \pt \rangle$ & 
\multicolumn{1}{c||}{$\dspt$} & $\langle \pt \rangle$ &
\multicolumn{1}{c|}{$\dspt$} \\ 
 {[GeV]} &  {[GeV]} & 
\multicolumn{1}{c||}{[pb/GeV]} &  [GeV] &
\multicolumn{1}{c|}{[pb/GeV]} \\ \hline
 0.12--0.28 &  0.20    & (1.38\,$\pm$\,0.00\,$\pm$\,0.04)$\times 10^{ 4}$ &  0.20   & (6.71\,$\pm$\,0.02\,$\pm$\,0.22)$\times 10^{ 3}$\\
 0.28--0.44 &  0.35    & (1.05\,$\pm$\,0.00\,$\pm$\,0.03)$\times 10^{ 4}$ &  0.35   & (5.11\,$\pm$\,0.01\,$\pm$\,0.17)$\times 10^{ 3}$\\
 0.44--0.60 &  0.51    & (5.85\,$\pm$\,0.01\,$\pm$\,0.15)$\times 10^{ 3}$ &  0.51   & (2.85\,$\pm$\,0.01\,$\pm$\,0.09)$\times 10^{ 3}$\\
 0.60--0.80 &  0.69    & (2.80\,$\pm$\,0.01\,$\pm$\,0.07)$\times 10^{ 3}$ &  0.69   & (1.37\,$\pm$\,0.01\,$\pm$\,0.04)$\times 10^{ 3}$\\
 0.80--1.00 &  0.89    & (1.20\,$\pm$\,0.01\,$\pm$\,0.03)$\times 10^{ 3}$ &  0.89   & (6.04\,$\pm$\,0.04\,$\pm$\,0.19)$\times 10^{ 2}$\\
 1.00--1.20 &  1.09    & (5.55\,$\pm$\,0.04\,$\pm$\,0.14)$\times 10^{ 2}$ &  1.09   & (2.83\,$\pm$\,0.03\,$\pm$\,0.08)$\times 10^{ 2}$\\
 1.20--1.40 &  1.29    & (2.75\,$\pm$\,0.03\,$\pm$\,0.07)$\times 10^{ 2}$ &  1.29   & (1.43\,$\pm$\,0.02\,$\pm$\,0.04)$\times 10^{ 2}$\\
 1.40--1.60 &  1.49    & (1.49\,$\pm$\,0.02\,$\pm$\,0.04)$\times 10^{ 2}$ &  1.49   & (7.88\,$\pm$\,0.14\,$\pm$\,0.20)$\times 10^{ 1}$\\
 1.60--1.80 &  1.69    & (8.51\,$\pm$\,0.14\,$\pm$\,0.24)$\times 10^{ 1}$ &  1.69   & (4.58\,$\pm$\,0.10\,$\pm$\,0.12)$\times 10^{ 1}$\\
 1.80--2.00 &  1.89    & (5.04\,$\pm$\,0.11\,$\pm$\,0.15)$\times 10^{ 1}$ &  1.89   & (2.77\,$\pm$\,0.08\,$\pm$\,0.08)$\times 10^{ 1}$\\
 2.00--2.20 &  2.09    & (3.25\,$\pm$\,0.09\,$\pm$\,0.10)$\times 10^{ 1}$ &  2.09   & (1.80\,$\pm$\,0.06\,$\pm$\,0.06)$\times 10^{ 1}$\\
 2.20--2.40 &  2.29    & (2.10\,$\pm$\,0.07\,$\pm$\,0.07)$\times 10^{ 1}$ &  2.29   & (1.15\,$\pm$\,0.05\,$\pm$\,0.04)$\times 10^{ 1}$\\
 2.40--2.60 &  2.50    & (1.41\,$\pm$\,0.05\,$\pm$\,0.05)$\times 10^{ 1}$ &  2.50   & (7.90\,$\pm$\,0.41\,$\pm$\,0.33)$\times 10^{ 0}$\\
 2.60--2.80 &  2.70    & (9.75\,$\pm$\,0.45\,$\pm$\,0.39)$\times 10^{ 0}$ &  2.70   & (5.43\,$\pm$\,0.33\,$\pm$\,0.23)$\times 10^{ 0}$\\
 2.80--3.00 &  2.90    & (7.49\,$\pm$\,0.40\,$\pm$\,0.29)$\times 10^{ 0}$ &  2.90   & (4.19\,$\pm$\,0.30\,$\pm$\,0.18)$\times 10^{ 0}$\\
 3.00--3.50 &  3.23    & (4.72\,$\pm$\,0.20\,$\pm$\,0.18)$\times 10^{ 0}$ &  3.23   & (2.79\,$\pm$\,0.16\,$\pm$\,0.11)$\times 10^{ 0}$\\
 3.50--4.00 &  3.73    & (2.54\,$\pm$\,0.15\,$\pm$\,0.10)$\times 10^{ 0}$ &  3.73   & (1.53\,$\pm$\,0.12\,$\pm$\,0.07)$\times 10^{ 0}$\\
 4.00--5.00 &  4.43    & (1.19\,$\pm$\,0.07\,$\pm$\,0.06)$\times 10^{ 0}$ &  4.43   & (7.03\,$\pm$\,0.56\,$\pm$\,0.51)$\times 10^{-1}$\\
 5.00--6.00 &  5.43    & (5.99\,$\pm$\,0.68\,$\pm$\,0.38)$\times 10^{-1}$ &  5.43   & (3.66\,$\pm$\,0.58\,$\pm$\,0.32)$\times 10^{-1}$\\
 6.00--8.00 &  6.80    & (2.85\,$\pm$\,0.36\,$\pm$\,0.21)$\times 10^{-1}$ &  6.83   & (1.99\,$\pm$\,0.33\,$\pm$\,0.19)$\times 10^{-1}$\\
 8.00--15.00 & 10.11    & (5.27\,$\pm$\,0.93\,$\pm$\,0.40)$\times 10^{-2}$ & 10.25   & (4.28\,$\pm$\,0.91\,$\pm$\,0.40)$\times 10^{-2}$\\
15.00--25.00 & 18.16\,$\pm$\,0.03    & (1.78\,$\pm$\,1.32\,$\pm$\,0.14)$\times 10^{-2}$ & 18.14\,$\pm$\,0.03   & (1.80\,$\pm$\,0.78\,$\pm$\,0.17)$\times 10^{-2}$\\
\hline
\end{tabular}
\caption{Differential inclusive charged pion production cross-sections
  $\dspt$ for $|\eta|<1.0$ and in the ranges $W>30$~GeV and
  $W>50$~GeV.  The first uncertainty is the statistical uncertainty
  and the second uncertainty is the systematic uncertainty. No value
  is given if the error on $\langle \pt \rangle$ is less than 0.01.}
\label{tab-ptl3b}
\end{center}
\end{table}


\end{document}